\pdfoutput=1
\documentclass[aps,prr,amsmath,amssymb,groupedaddress,twocolumn]{revtex4-2}
\usepackage{bm}
\usepackage{bbold}
\usepackage{graphicx}
\usepackage[colorlinks=True]{hyperref}

\begin{document}

\preprint{LA-UR-23-29443}

\title{Relativistic materials from an alternative viewpoint}


\author{Ann E. Mattsson}
\email[]{aematts@lanl.gov}
\author{Daniel A. Rehn}
\email[]{rehnd@lanl.gov}
\affiliation{Computational Physics Division, Los Alamos National Laboratory, Los Alamos, NM 87545}


\date{\today}
\begin{abstract}
Electrons in materials containing heavy elements are fundamentally
relativistic and should in principle be described using the Dirac
equation. However, the current standard for treatment of electrons in
such materials involves density functional theory methods originally
formulated from the Schr\"{o}dinger equation.  While some extensions
of the Schr\"{o}dinger-based formulation have been explored, such as
the scalar relativistic approximation with or without spin-orbit
coupling, these solutions do not provide a way to fully account for
all relativistic effects of electrons, and the language used to
describe such solutions are still based in the language of the
Schr\"{o}dinger equation. In this article, we provide a different
method for translating between the Dirac and Schr\"{o}dinger
viewpoints in the context of a Coulomb potential. By retaining the
Dirac four-vector notation and terminology in taking the
non-relativistic limit, we see a much deeper connection between the
Dirac and Schr\"{o}dinger equation solutions that allow us to more
directly compare the effects of relativity in the angular and radial
functions. Through this viewpoint, we introduce the concepts of
\emph{densitals} and \emph{Dirac spherical harmonics} that allow us to
translate more easily between the Dirac and Schr\"{o}dinger
solutions. These concepts allow us to establish a useful language for
discussing relativistic effects in materials containing elements
throughout the full periodic table and thereby enable a more
fundamental understanding of the effects of relativity on electronic
structure.
\end{abstract}


\maketitle

\section{Introduction\label{sec:intro}}

Solid state materials containing heavy elements are currently studied
using computational methods based on the Schr\"{o}dinger equation
(SE).  Among the most popular approaches is density functional theory
(DFT),~\cite{HK,KS} which was originally developed as a reformulation
of the Schr\"{o}dinger equation.  For heavy elements, however,
relativistic effects are known to play an essential role in
determining materials properties.  While some extensions of SE-based
DFT such as the scalar relativstic (SR)~\cite{koelling} and SR with
spin-orbit coupling (SR+SO) are frequently used, these methods do not
include all relativistic effects and introduce approximations that so
far have not been adequately assessed. On the other hand, the
relativistic formulation of DFT based on the Dirac equation was
developed nearly 40 years ago in the work of Rajagopal and
Galloway~\cite{rajagopal1973inhomogeneous,rajagopal1978inhomogeneous}, and also by MacDonald and Vosko.~\cite{macdonald1979relativistic} This relativistic
formulation, which we refer to as RDFT, has not been used or
implemented extensively in the context of solid state materials,
though some codes have done so using various methodologies.~\cite{rehn2020dirac,rehn2020relativisitc,pashov2020questaal,sprkkr,ebert2011calculating,eschrig2004relativistic,kadek2019all,liu2003beijing,zhao2021quasi} On the other
hand, the quantum chemistry community frequently uses methods based on
the Dirac equation for the study of heavy elements.~\cite{DIRAC18,fritzsche2002relci,fischer2007mchf,visscher1994relativistic,belpassi2008all,belpassi2011recent,shiozaki2018bagel,quiney2002relativistic,willetts2000magic}

In order to advance our predictive capabilities for materials
containing heavy elements and assess the SR and SR+SO approximations
in the context of solids, we need to move towards the RDFT formulation
and solve the corresponding Dirac-Kohn-Sham equations in order to
establish a baseline from which to assess the approximations made with
existing methods. However, to do this, we must also address the fact
that a significant language barrier exists in moving between the Dirac
and Schr\"{o}dinger solutions. This language barrier originates from
the way we refer to states in the Dirac and SE solutions to the
hydrogen-like atom, i.e., an isolated ion (nucleus) with nuclear
charge $Z$ and one electron.  The SE solutions give us a quantum
number $n$ describing the energy, and quantum numbers $l$ and $m$
giving the orbital angular momentum, as well as a spin index
$s=\uparrow,\downarrow$. On the other hand, the Dirac equation applied
to the same problem gives us a different set of quantum numbers, $N$,
$\kappa$, $j$, and $j_3$, defined below.  While $l$ also appears in
the Dirac equation, it depends on the value of $\kappa$ or $j$ and is not a
good quantum number. In addition, the spin index in the Dirac
solutions arises naturally in the Dirac four-vector notation, along
with the two positron spin solutions. These fundamental differences in
quantum numbers make it important to develop a language and methodology for
translating between the SE and Dirac pictures.

The primary focus of this article is to establish a new language for
translating between the Dirac and SE solutions. We do this by focusing
on the hydrogen-like atom, deriving analytic solutions to the problem
with the Dirac equation, and then taking the non-relativistic limit in
a particular way that allows the SE solutions to be understood in the
context of the four-vector solutions to the Dirac equation. In doing
so, we define two concepts that establish a language for translating:
\emph{densitals} and \emph{Dirac spherical harmonics}.  We then use
these concepts to analyze states in the SE and Dirac solutions.  These
concepts also allow us to translate from the traditional SE
terminology (e.g., $s$, $p$, $d$, $f$ states) to the more general
Dirac terminology we define below.  The benefit of this approach is
that it enables a much more direct and informative comparison of the
effects of relativity on any orbital we wish to consider. Furthermore,
it allows us to visualize the effect of relativity on contraction of
orbitals or radial functions in a direct way, and also sheds light on
why, in some cases, DFT$+U$ methods may not be adequate to account for
all types of orbital contraction, which is known to be important in
the study of lanthanides and actinides. 

A further benefit of developing this language and methodology is that
it will enable us to discuss magnetic properties of systems in a more
natural way.  The RDFT formulation tells us that the ground state
energy of a system should depend not only on the static electron
density $n$ through the effective scalar potential $v_\mathrm{eff}$,
but also the current density $\pmb{J}$, through the effective
vector potential $\mathbf{A}_\mathrm{eff}$. This lends itself to a
more general way to discuss magnetism in the context of solids than
when using the SR and SR+SO methodologies, in which the spin and
orbital contributions to the total current are decoupled and the
orbital contribution is usually only included when adding the
so-called orbital polarization (OP) term.~\cite{eriksson1990orbital1,eriksson1990orbital2,soderlind2007delta,soderlind2008quantifying}
These corrections introduce approximations that are difficult to
assess without a full solution to the Dirac-Kohn-Sham
equations.

We point out that most of the material presented in this article has been published and discussed before, in fact, much of it is textbook material. The key to this work is to put the pieces together in a way that can enhance our understanding and facilitate discussions. We do not pretend that our way is the only clarifying way the pieces can be put together, but it is a way that has worked well for us. We also expect the language to evolve as others adopt it.

Fundamentally, the purpose of the present work is to provide a
framework for discussing relativistic states in materials, one that at
the same time provides a way to translate from the commonly-used SE
concepts to the more general Dirac concepts. To do this, we proceed as
follows: in Sec.~\ref{sec:dirac} we define the Dirac equation in full
four-vector notation. In Sec.~\ref{sec:hydrogen} we solve the Dirac
equation for the hydrogen-like atom and take the non-relativistic
limit in a manner that is different from most textbook treatments. In
Sec.~\ref{sec:densitals} we introduce the concept of \emph{densitals}
and show comparisons of SE and Dirac solutions using this concept.
We follow with a discussion of how this compares to the SR
approximation in Sec.~\ref{sec:discussion} and conclude with ideas for
future work that use this framework~\ref{sec:future}.

\section{The Dirac Equation\label{sec:dirac}}
The Dirac equation in $4\times4$ matrix form is
\begin{equation}
  \left(c\,\pmb{\alpha}\cdot \mathbf{p} +
  \mathbb{1} V(\mathbf{r}) + \beta mc^2\right)
  \psi_i(\mathbf{r}) = \varepsilon_i \psi_i(\mathbf{r}) \, ,
  \label{eq:ch2:diracham3}
\end{equation}
where the canonical momentum is $\mathbf{p}=-i \hbar \mathbf{\nabla}$, the wave function is a $4-$component spinor
\begin{equation}
  \psi_i = \begin{pmatrix}\psi_{1,i} \\\psi_{2,i} \\\psi_{3,i} \\\psi_{4,i} \end{pmatrix}
  = \begin{pmatrix}\psi_{A,i} \\\psi_{B,i} \end{pmatrix} \, ,
\end{equation}
where $\psi_{A,i}\; (\psi_{B,i})$ are the two upper (lower) components, and the matrices are
\begin{equation}
  \alpha_k = \begin{pmatrix}0 & \sigma_k \\ \sigma_k & 0\end{pmatrix},
    \;
    \beta = \begin{pmatrix}I & 0 \\ 0 & -I\end{pmatrix},
      \;
      I = \begin{pmatrix}1 & 0 \\ 0 & 1 \end{pmatrix},
      \;
      \mathbb{1} = \begin{pmatrix}I & 0 \\ 0 & I \end{pmatrix},
\end{equation}
where $\sigma_k$ are the Pauli matrices.

\section{The hydrogen-like atom: Dirac solution\label{sec:hydrogen}}
The electrons in a hydrogen-like atom with nuclear charge $Z$ move in
a Coulomb potential defined by
\begin{equation}
V(\mathbf{r}) = V(r) = -\frac{Z e_c^2}{4 \pi r} \, ,
\label{eq:v}
\end{equation}
where $e_c$ is the electron charge.

The solution to the Dirac equation in the presence of a Coulomb
potential is described in the text book \emph{Advanced Quantum Mechanics}
by Sakurai~\cite{sakurai1967advanced}, and we follow most of the
notation given therein. Following this, we define the electron charge,
$e_c$, as a negative quantity, $e_c=-|e_c|$.

The key concept we use from Ref.~\cite{sakurai1967advanced} is that we
transform the radial differential equations to ordinary equations
using an \emph{ansatz} containing a few parameters and expansion
coefficients in the radial coordinate that are subsequently
determined so that the differential equation is fulfilled. A major
difference, however, is that we do the derivation from the second
order radial differential equation for only the upper components and
derive the lower components from those solutions. This is because we
wish to obtain a direct comparison with the second order radial Schr\"odinger equation.

The energy eigenvalues for the Dirac equation with the potential as in
Eqn.~\ref{eq:v} are
\begin{equation}
e_{N \kappa}=\frac{\varepsilon_{N \kappa}}{ m c^2 } =\sqrt{\frac{(N+\sqrt{\kappa^2-\gamma^2})^2}{(N+\sqrt{\kappa^2-\gamma^2})^2+\gamma^2}} \, , \label{eqn:e}
\end{equation} 
where $N \geq 0$ is an integer, $\kappa \neq 0$ is an integer, and
\begin{equation}
\gamma = \frac{Z e_c^2}{4 \pi \hbar c}=Z \alpha
\end{equation}
with $\alpha \approx 1/137$ the fine-structure constant.  The
corresponding eigenfunctions are
\begin{equation}
 \psi^{\kappa<0}_{N j j_3}= \begin{pmatrix} g_{N \kappa}(r) \sqrt{\frac{1}{2}(1+\frac{j_3}{j})} &Y_{j-\frac{1}{2}}^{j_3-\frac{1}{2}}(\theta,\phi) \\
g_{N \kappa}(r) \sqrt{\frac{1}{2}(1-\frac{j_3}{j})} &Y_{j-\frac{1}{2}}^{j_3+\frac{1}{2}}(\theta,\phi) \\
-i f_{N \kappa}(r) \sqrt{\frac{1}{2}(1-\frac{j_3}{j+1})} &Y_{j+\frac{1}{2}}^{j_3-\frac{1}{2}}(\theta,\phi)\\
i f_{N \kappa}(r) \sqrt{\frac{1}{2}(1+\frac{j_3}{j+1})} &Y_{j+\frac{1}{2}}^{j_3+\frac{1}{2}}(\theta,\phi) \label{eqn:wfneg}
\end{pmatrix},
\end{equation}
when $\kappa < 0$ and $j=-\kappa-\frac{1}{2}$,
and  
\begin{equation}
 \psi^{\kappa>0}_{N j j_3}= \begin{pmatrix} -g_{N \kappa}(r) \sqrt{\frac{1}{2}(1-\frac{j_3}{j+1})} &Y_{j+\frac{1}{2}}^{j_3-\frac{1}{2}}(\theta,\phi) \\
g_{N \kappa}(r) \sqrt{\frac{1}{2}(1+\frac{j_3}{j+1})} &Y_{j+\frac{1}{2}}^{j_3+\frac{1}{2}}(\theta,\phi) \\
i f_{N \kappa}(r) \sqrt{\frac{1}{2}(1+\frac{j_3}{j})} &Y_{j-\frac{1}{2}}^{j_3-\frac{1}{2}}(\theta,\phi)\\
i f_{N \kappa}(r) \sqrt{\frac{1}{2}(1-\frac{j_3}{j})} &Y_{j-\frac{1}{2}}^{j_3+\frac{1}{2}}(\theta,\phi) \label{eqn:wfpos}
\end{pmatrix},
\end{equation}
when $\kappa > 0$ and  $j=\kappa-\frac{1}{2}$.\\

Here $j_3$ is the projection of $j$ so that $-j \leq j_3 \leq j$ in
integer steps, and $Y_l^m(\theta,\phi)$ are (complex) Laplace
spherical harmonics. The upper and lower radial functions are
\begin{eqnarray}
 g_{N \kappa}(r) &=& \left( \frac{\sqrt{1-e_{N \kappa}^2}}{\gamma} \frac{Z}{a_{bohr}} \right)^{3/2} N_0 \, e^{-\rho_D} \rho_D^{\, p-1} \times \nonumber\\
                           && \hspace{3cm} \times \sum_{m=0}^N b_m \rho_D^m  \, , \label{eqn:gNkappa} \\
 f_{N \kappa}(r) &=&  \left( \frac{\sqrt{1-e_{N \kappa}^2}}{\gamma} \frac{Z}{a_{bohr}} \right)^{3/2} N_0 \, e^{-\rho_D} \rho_D^{\, p-1} \times \nonumber \\
                            && \hspace{3cm} \times \sum_{m=0}^N a_m \rho_D^m \, , \label{eqn:fNkappa}
 \end{eqnarray}
 where $\rho_D$ is dimensionless,
\begin{equation}
\rho_D = \frac{\sqrt{1-e_{N \kappa}^2}}{\gamma} \frac{Z}{a_{bohr}} \, r \, , \label{eqn:rho}
\end{equation}
as is $p^2= \kappa^2-\gamma^2$. The coefficients in the sums are obtained by the recursion relations
\begin{equation}
 b_1 = \frac{\left[\sqrt{\frac{1+e_{N \kappa}}{1-e_{N \kappa}}} \frac{1}{\gamma} ( p+\kappa)- (2 N -1) \right] b_0}{(2p+1)} \, , 
 \end{equation}
 \begin{widetext}
 \begin{equation}
 b_{q+2} = \frac{\left[ \sqrt{\frac{1+e_{N \kappa}}{1-e_{N \kappa}}} \frac{1}{\gamma} \left( p+\kappa-(q+1)(q+2p)\right) - \left(2 (N-(q+1))-1 \right)\right]b_{q+1} -\sqrt{\frac{1+e_{N \kappa}}{1-e_{N \kappa}}} \frac{1}{\gamma} 2 (N-q) b_{q}}{  (q+2)(q+2+2 p)}  \, , \label{eqn:Diracbcoeff}
\end{equation}	
\end{widetext}
and
\begin{eqnarray}
\gamma a_0 &=& (p+\kappa) b_0  \, ,\\
\sqrt{\frac{1+e_{N \kappa}}{1-e_{N \kappa}}} a_{q} + \gamma a_{q+1} &=& -b_q + \left((q+1+p)+\kappa\right) b_{q+1} \, . \nonumber \\
&& \label{eqn:acoeff}
\end{eqnarray}
The normalization constant is given by
\begin{equation}
N_0=\frac{1}{ \sqrt{ \sum_{t=0}^{2N} c_t \frac{\Gamma(1+2p+t)}{2^{1+2p+t}}}}  \label{eqn:norm}
\end{equation}
where $\Gamma$ is the gamma function and
\begin{equation}
 c_t = \left\{ \begin{matrix} {\sum\limits_{i=0}^t} &\left(a_{t-i} a_i + b_{t-i} b_i \right) \quad & 0 \leq t \leq N \\ \\  {\sum\limits_{i=t-N}^N} &\left(a_{t-i} a_i + b_{t-i} b_i \right) \quad & N \leq t \leq 2 N \end{matrix} \right. \, . \label{eqn:ct}
\end{equation}

\subsection{The non-relativistic limit}

The Schr\"odinger equation is the non-relativistic limit of the Dirac
equation, meaning $|V| \ll mc^2$. For the potential in Eqn.~\ref{eq:v}
this limit becomes
\begin{equation}
\frac{\sqrt{1-e_{N \kappa}^2} \gamma}{\rho_D} \ll 1 \, , \label{eqn:nrlimit}
\end{equation}
taking into account Eqn.~\ref{eqn:rho}.

In the Dirac equation the eigenenergies of a system includes the rest
mass ($mc^2$) but for a non-relativistic system the energy is instead
relative to the dominant energy contribution of the rest mass. This
means that $e_{NR} = e_{N \kappa}-1$ and that $0< - e_{NR} \ll 1$. Note
that this is consistent with the non-relativistic limit in
Eqn.~\ref{eqn:nrlimit}. Expanding the energy in Eqn.~\ref{eqn:e} in
terms of $\gamma$ we arrive at
\begin{equation}
e_{NR} = -\frac{1}{2} \frac{\gamma^2}{(N+|\kappa|)^2} \equiv  -\frac{1}{2} \frac{(Z \alpha)^2}{n^2} \label{eqn:eNR}
\end{equation}
and we can identify $N+|\kappa|$ with the non-relativistic principal
quantum number $n$, $n = 1,2,3,\ldots$.  We will soon also connect the
other two Dirac quantum numbers $j$ (or equivalently $\kappa$) and
$j_3$ to the non-relativistic quantum numbers $l$ and $m$.

It is straightforward to derive that
\begin{equation}
\sqrt{1-e_{N \kappa}^2} \stackrel{\gamma \rightarrow 0}{\longrightarrow} \frac{\gamma}{n} \, .
\end{equation}
This gives that the non-relativistic criteria in Eqn.~\ref{eqn:nrlimit} becomes
\begin{equation}
\frac{\gamma^2}{n \, \rho} \ll 1 \, ,
\end{equation}
with the dimensionless radial coordinate in Eqn.~\ref{eqn:rho} becoming
\begin{equation}
\rho = \frac{Z}{n \, a_{bohr}} \, r \, .
\end{equation}
The recursion relations for the coefficients $b^{NR}_m$ and $a^{NR}_m$,
and the value of $p_{NR}$ are determined from carefully eliminating
terms that are $\gamma^2/\rho$ smaller than other terms in the same
equations that the corresponding Dirac quantities are determined
from. The resulting equations are indeed the same equations we would
get if we put the ansatz directly into the Schr\"odinger equation.
The results are
\begin{equation}
p_{NR}(p_{NR}-1) = \kappa (\kappa +1) \, ,  \label{eqn:pNR}
\end{equation}
and
\begin{eqnarray}
b^{NR}_{q+1}&=&\frac{-2(n-p_{NR}-q)}{(q+1)(2p_{NR}+q)} b^{NR}_q\, , \label{eqn:SEbm} \\
a^{NR}_q &=& 0 \, ,
\end{eqnarray}
where we have used that $N+p_{NR} = n$, determined from identification
with the non-relativistic energy as above.

Equation~\ref{eqn:pNR} gives two solutions
\begin{equation}
 p_{NR} = \left\{ \begin{matrix}& -\kappa &= j-\frac{1}{2}+1 \ \ \ \ \ \kappa<0 \\ \\
                                                & \kappa + 1 &= j+\frac{1}{2}+1 \ \ \ \ \ \kappa>0 \end{matrix} \right. \label{eqn:SEp}
\end{equation}
Identifying $p_{NR}-1=l$, with $l$ the non-relativistic angular
momentum quantum number, we arrive at
\begin{equation}
g_{n l}(r) = \left( \frac{Z}{n \, a_{bohr}} \right)^{3/2} N_0 \, e^{-\rho} \rho^{\, l} \sum_{m=0}^{n-l-1} b_m \rho^m \, , \label{eqn:gnl} 
\end{equation}
which, with some good bookkeeping, can be reduced to the more
recognizable form
\begin{equation}
g_{n l} (r) =  \left( \frac{2 Z}{n \, a_{bohr}} \right)^{3/2} \sqrt{\frac{(n-l-1)!}{2 n (n+l)!}} e^{-\rho} (2 \rho)^{\, l} L_{n-l-1}^{(2l+1)}(2 \rho) \, ,
\label{eqn:gnlfinal}
\end{equation}
 where $L_a^{(b)}$ is the generalized Laguerre polynomial.
 
To connect the Dirac solutions even more to the non-relativistic limit, we replace
$m=j_3-1/2$ in both the negative and positive $\kappa$ wave functions
and arrive at
\begin{equation}
 \psi^{\kappa<0}_{nlm} = \begin{pmatrix} g_{nl}(r) \sqrt{\frac{(l+1)+m}{2l+1}} &Y_{l}^{m}(\theta,\phi) \\
g_{nl}(r) \sqrt{\frac{(l+1)-(m+1)}{2l+1}} &Y_{l}^{m+1}(\theta,\phi) \\
-i f_{nl}(r) \sqrt{\frac{(l+1)-m}{2(l+1)+1}}  &Y_{l+1}^{m}(\theta,\phi)\\
i f_{nl}(r) \sqrt{\frac{(l+1)+(m+1)}{2(l+1)+1}}  &Y_{l+1}^{m+1}(\theta,\phi) \label{eqn:lmwfneg}
\end{pmatrix},
\end{equation}
using that $l=j-\frac{1}{2};\; -\kappa=l+1; \; m=j_3-\frac{1}{2}$ for
$\kappa < 0$, and
\begin{equation}
 \psi^{\kappa>0}_{nlm} = \begin{pmatrix} -g_{nl}(r) \sqrt{\frac{l-m}{2l+1}}&Y_{l}^{m}(\theta,\phi) \\
g_{nl}(r) \sqrt{\frac{l+(m+1)}{2l+1}} &Y_{l}^{m+1}(\theta,\phi) \\
i f_{nl}(r) \sqrt{\frac{l+m}{2(l-1)+1}} &Y_{l-1}^{m}(\theta,\phi)\\
i f_{nl}(r) \sqrt{\frac{l-(m+1)}{2(l-1)+1}} &Y_{l-1}^{m+1}(\theta,\phi) \label{eqn:lmwfpos}
\end{pmatrix},
\end{equation}
using that $l=j+\frac{1}{2};\; \kappa=l; \; m=j_3-\frac{1}{2}$ for
$\kappa > 0$.  If we now set $f_{nl}(r)=0$, as we derived above for
the SE, we can add and subtract the two wave functions with suitable
coefficients to arrive at the fact that both the upper spin up, $\psi_{1,i}$,
and spin down, $\psi_{2,i}$, components can be described by
\begin{equation}
\psi_{nlm}=g_{nl}(r) Y_l^m(\theta,\phi) \, .
\end{equation}
The spin and angular momentum are indeed decoupled in the
non-relativistic limit, and both spin and angular momentum are now
good quantum numbers.

\section{Densitals\label{sec:densitals}}
SE solutions are often plotted using only the angular part of the wavefunctions,
not the radial part. Most often this is done using the real, as opposed to complex (or Laplace),
spherical harmonics, and these are typically referred to as orbitals. However, the
Dirac solution in Eqns.~\ref{eqn:wfneg} and~\ref{eqn:wfpos} cannot be
illustrated in this manner due to the fact that we have a mixture of different angular momenta functions in each component, as well as different radial functions in the upper and lower components. Because of this, we will introduce the the concept of \emph{densitals}~\cite{densitals} for illustrating our solutions.

While the wavefunctions or orbitals of a system are not observables, the electron density formed by these orbitals
\begin{equation}
\rho_e=\sum_i f_i |\psi_i|^2 
\label{eqn:dens}
\end{equation}
with $f_i$ an occupancy factor.
The density is also the primary quantity in the different computational methods based on density functional theory~\cite{HK,KS}, that are prevalent in materials physics and many other fields. \emph{Densitals} are illustrative objects based on the partial change densities via plotting of the volume where their probability density is larger than a cutoff value,
\begin{equation}
|\psi_i|^2 > \rho_{\rm cutoff} \, .
\end{equation}

\begin{figure*}[!ht]
\includegraphics[width=13cm]{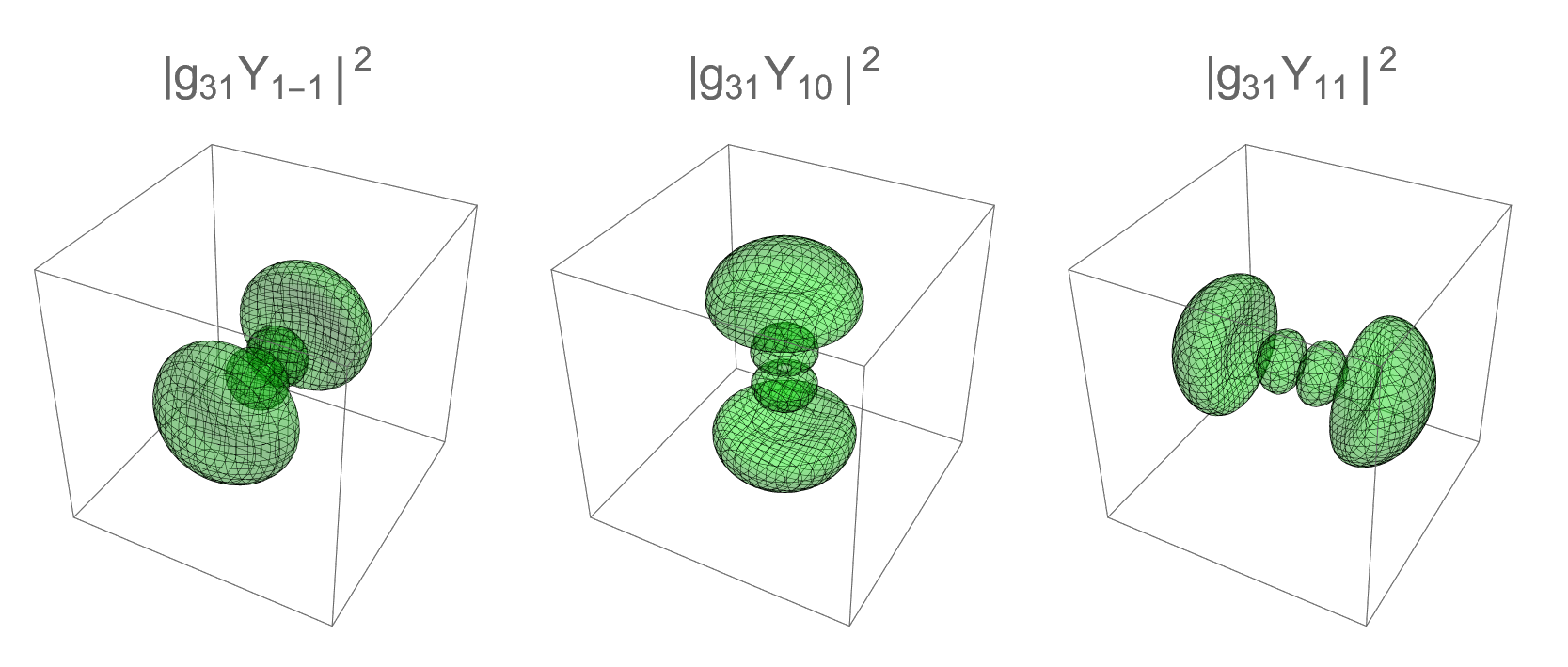}
\caption{\label{fig:realdensitals}Densitals for the third shell $p$-states based on real spherical harmonics. Each state can accommodate one spin-up and one spin-down electron. We have $6$ $p$-states.}
\end{figure*}
In Fig.~\ref{fig:realdensitals} we show the densitals for the $3$ different SE $p$-states ($n=3, l=1, m=-1,0,1$) using the usual real spherical harmonics and the radial function in Eqn.~\ref{eqn:gnlfinal}.

\begin{figure*} [!ht]
\includegraphics[width=13cm]{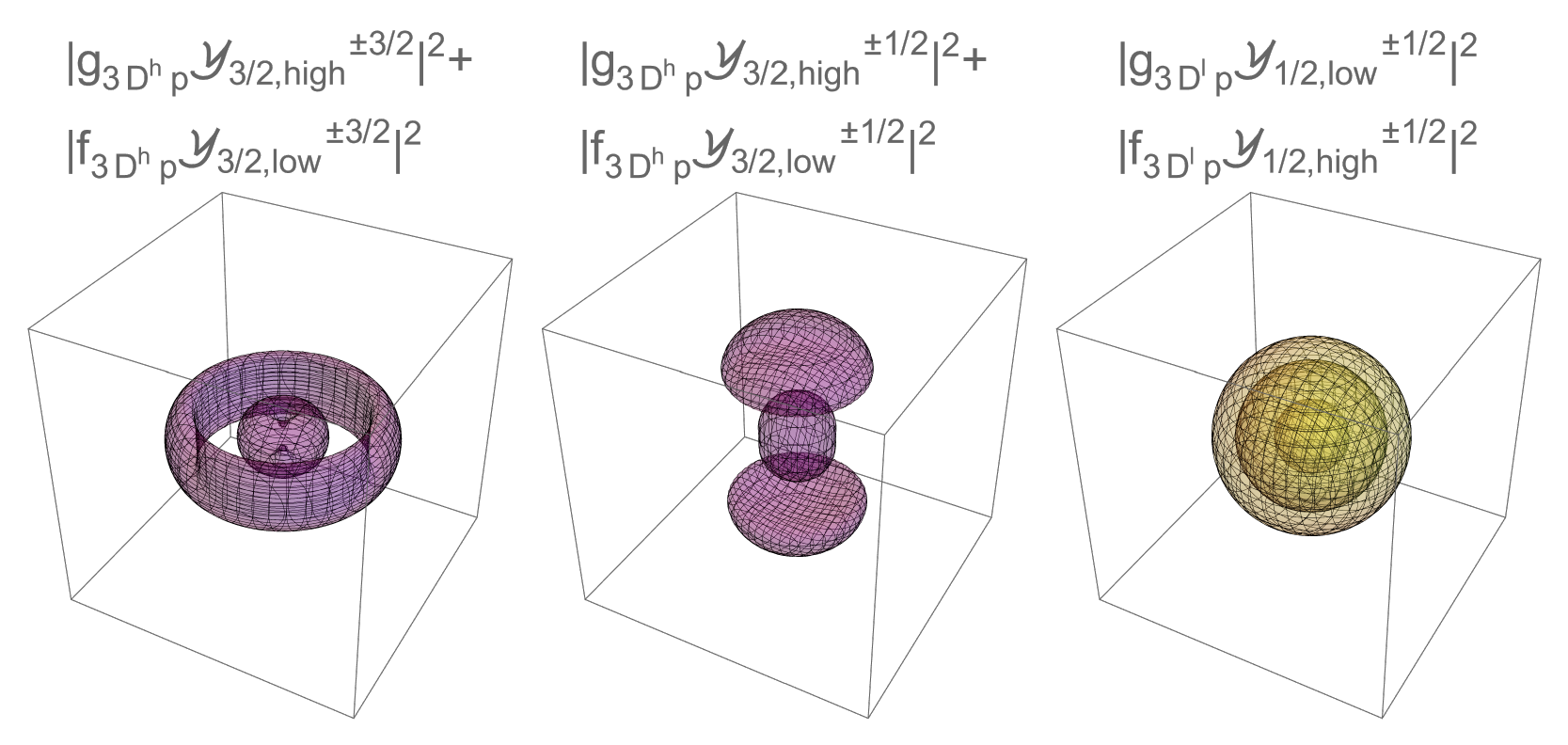}
\caption{\label{fig:DDiracdensitals}Densitals based on the Dirac spherical harmonics for the $3$ distinct Dirac equation solutions in Eqns.~\ref{eqn:wfneg} and~\ref{eqn:wfpos} corresponding to the $3p$-densitals in Fig.~\ref{fig:realdensitals}. Dirac $high$ densitals are in purple and the Dirac $low$ densital is in yellow.}
\end{figure*}

However, as mentioned above we cannot use the real spherical harmonics for the Dirac solution in Eqns.~\ref{eqn:wfneg} and~\ref{eqn:wfpos}. We therefore define \emph{Dirac spherical harmonics}~\cite{Diracsphericalharmonics} as
\begin{widetext}
\begin{eqnarray}
\mathcal{Y}_{j,high}^{j_3} &=& \sqrt{\frac{(l+1)+m}{2 l + 1}} Y_l^m \begin{pmatrix} 1\\0 \end{pmatrix} + \sqrt{\frac{(l+1)-(m+1)}{2 l + 1}} Y_l^{m+1} \begin{pmatrix} 0\\1 \end{pmatrix} \label{eqn:DiracYhigh}\\
\mathcal{Y}_{j,low}^{j_3} &=& - \sqrt{\frac{l-m}{2 l + 1}} Y_l^m \begin{pmatrix} 1\\0 \end{pmatrix} + \sqrt{\frac{l+(m+1)}{2 l + 1}} Y_l^{m+1} \begin{pmatrix} 0\\1 \end{pmatrix} \label{eqn:DiracYlow}
\end{eqnarray}
\end{widetext}
where $j=l+\frac{1}{2}$ for $high$ and $j=l-\frac{1}{2}$ for $low$ and $j_3 = m \pm \frac{1}{2}$ so that $-j, -j+1, \ldots \leq j_3 \leq \ldots, j-1,j$. From Eqns.~\ref{eqn:lmwfneg} and~\ref{eqn:lmwfpos} we have that the upper components are described by these Dirac spherical harmonics. The lower components come from the same formulas but the $\kappa<0$ solution uses the $low$ formula with $l$ replaced by $l+1$ and the $\kappa>0$ solution uses the $high$ formula with $l$ replaced by $l-1$. This shift in $l$ values for the lower components means that the upper and lower components will have the same values for $j$ and $j_3$. We see that the upper and lower components separately use a linear combination of $Y_l^m$ with $l$ fixed and different $m$. We also note that we already in the separate upper and lower components are mixing spin and the projection of the angular momentum. The angular momentum itself gets mixed when upper and lower components are taken into account together, as when forming the density or densitals. We should note that the Dirac spherical harmonics are eigenfunctions of the spin-orbit coupling operator. In Fig.~\ref{fig:DDiracdensitals} we show the densitals for the $3$ distinct Dirac solutions for $n=3, l=1$, that is, two Dirac high states, $3D^h p_{\pm 1/2}$ and $3D^h p_{\pm 3/2}$, with $j=3/2$ and the single Dirac low state, $3D^l p_{\pm 1/2}$, with $j=1/2$. The notation used here is $n \, D^{h(igh),l(ow)}l_{j_3}$, with $n$ the non-relativistic principal quantum number, $l$ the non-relativistic angular momentum, $j_3$, which is often omitted, the projection of the relativistic total angular momentum $j$, with $j=l+1/2$ for $D^h$ and $j=l-1/2$ for $D^l$.

\begin{figure*} [!ht]
\includegraphics[width=13cm]{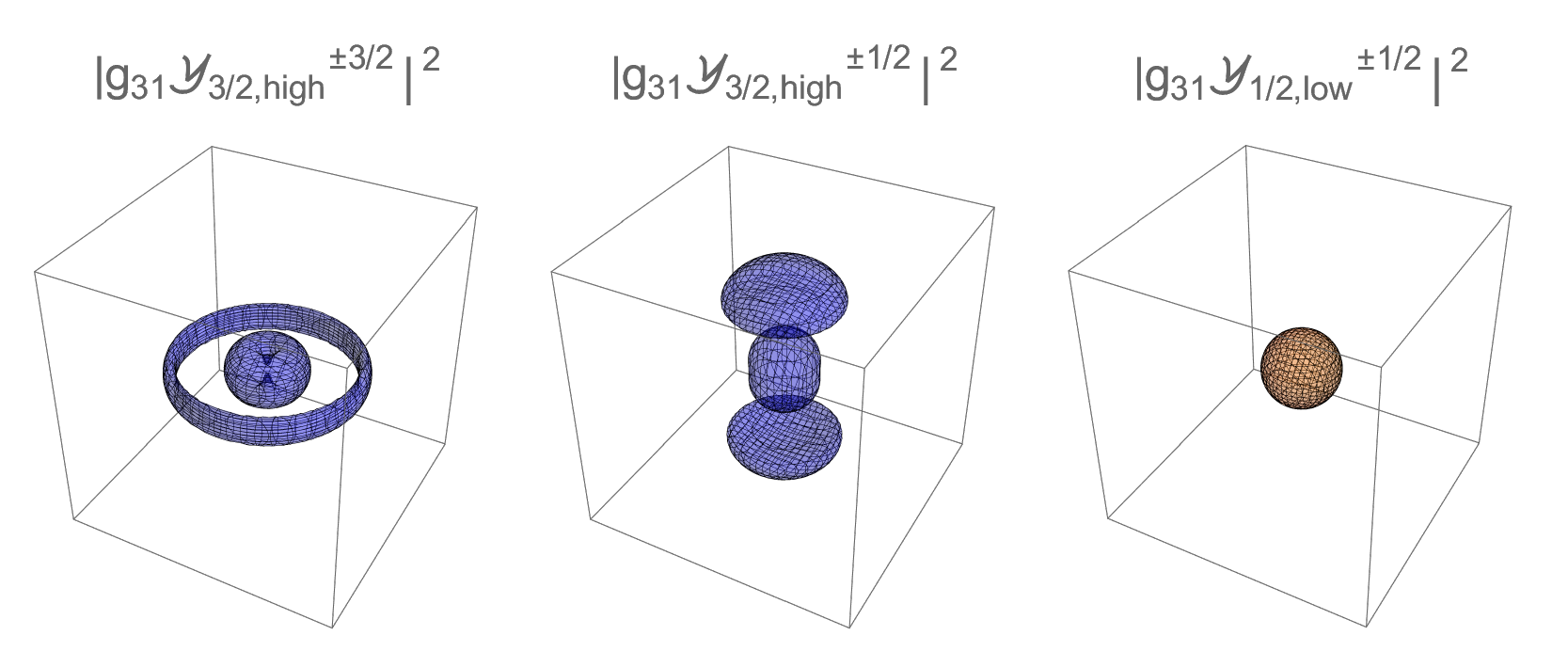}
\caption{\label{fig:Diracdensitals}Densitals based on the Dirac spherical harmonics for the $3p$ SE equation solutions in Eqns.~\ref{eqn:lmwfneg} and~\ref{eqn:lmwfpos}. These densitals represents the same solution as the $3p$-densitals in Fig.~\ref{fig:realdensitals}. Dirac $high$ densitals are in blue and the Dirac $low$ densital is in orange.}
\end{figure*}

A comparison of Figs.~\ref{fig:realdensitals} and \ref{fig:DDiracdensitals} does not allow for an easy interpretation or translation between the Dirac and SE solutions. However, if we plot the densitals for the SE solution using the Dirac spherical harmonics, as in Eqns.~\ref{eqn:lmwfneg} and~\ref{eqn:lmwfpos} (with $f_{nl}(r)=0$) as in Fig.~\ref{fig:Diracdensitals}, we see a clear correspondence between the Dirac and the SE solutions.
The correspondence between the SE and Dirac solutions when seen from the viewpoint of Dirac spherical harmonics is the main message of this work. We should be able to better understand lanthanide and actinide materials from SE calculations using this alternative viewpoint.

\section{Discussion\label{sec:discussion}}
\subsection{The two Dirac solutions}
The two solutions can be written as a single solution if we allow the use of spherical harmonics  $Y_\kappa^{m_\kappa}$ with negative $\kappa$. We have, however, chosen to use the definition $Y_\kappa^{m_\kappa}=Y_{-\kappa-1}^{m_\kappa}$ for negative $\kappa$, to more easily relate to standard SE language with $l$ being a positive integer.
We thus have two types of Dirac solutions, $high$ and $low$ or $\kappa<0$ and $\kappa>0$. We have chosen the names $high$ and $low$ for several reasons. We will discuss this when considering the translations from $l$ to $\kappa, j$. Since $l$ and $\kappa$ have different relations in the $high$ and $low$ solutions, Dirac solutions with the same $l$ can have different energies because of the different values of $\kappa^2$, see Eqn.~\ref{eqn:e}, and they will have different radial functions, see Eqn.~\ref{eqn:Diracbcoeff}. In fact, the $high$ solution will have $|\kappa|=(j+1/2)=l+1$ and the $low$ solution $|\kappa|=(j+1/2)=l$, and the $high$ solution will have a higher energy than the $low$ solution. Of course this is the spin-orbit splitting. The $high$ solution will also have a higher value of $j=l+1/2$ than the $low$ solution, $j=l-1/2$, for a fixed $l$. Note that in our notation the value of $j$ is indicated by the labels $high$ and $low$ while an index is giving the value of $j_3$. We find this notation easier to use in discussions than the standard notation since it is more easily related to energy and how spin and angular momentum are related.
In the non-relativistic limit the Dirac solution in Eqns.~\ref{eqn:lmwfneg} and~\ref{eqn:lmwfpos} (with $f_{nl}(r)=0$), corresponds to spin-up and spin-down solutions along the angular momentum $z$-axis, or as it is commonly phrased, angular momentum and spin are parallel or anti-parallel. When the spin and angular momentum couples, the total angular momentum $j$ will be $high$ ($j=l+1/2$) for the parallel solution and $low$ ($j=l-1/2$) for the anti-parallel solution. 

The main confusion that can still arise is when comparing states where $l$ is not the same. Of course, this is because the fact that we are not used to translating between the relativistic $\kappa, j$ and the non-relativistic $l$. For example, $nD^l p$ and $nD^h s$ will be degenerate in energy in the Dirac solution since they have the same $|\kappa|$. Indeed, evaluating $|\mathcal{Y}_{l+1/2,low}^{j_3}|^2$ and $|\mathcal{Y}_{l+1/2,high}^{j_3}|^2$ shows that they are the same function, as they should since they have the same $j$ and $j_3$ values, and the densitals only differ by their different radial functions which are dependent also on the sign of $\kappa$. 

\subsection{Main effect of relativity}
There are many more or less fundamental ways of understanding that the main effect of relativity is to contract the electron density closer to the nuclei. In the planetary view of electrons and nuclei, we know that a planet that moves faster around its sun also needs to contract its orbit since the area swept per unit time needs to be constant. In the positron/electron picture we understand that the mingling of positrons and electrons in the deeper parts of the potential close to the nucleus can allow for more electrons to occupy low energy states in this region. 

\begin{figure} [!ht]
\includegraphics[width=6cm]{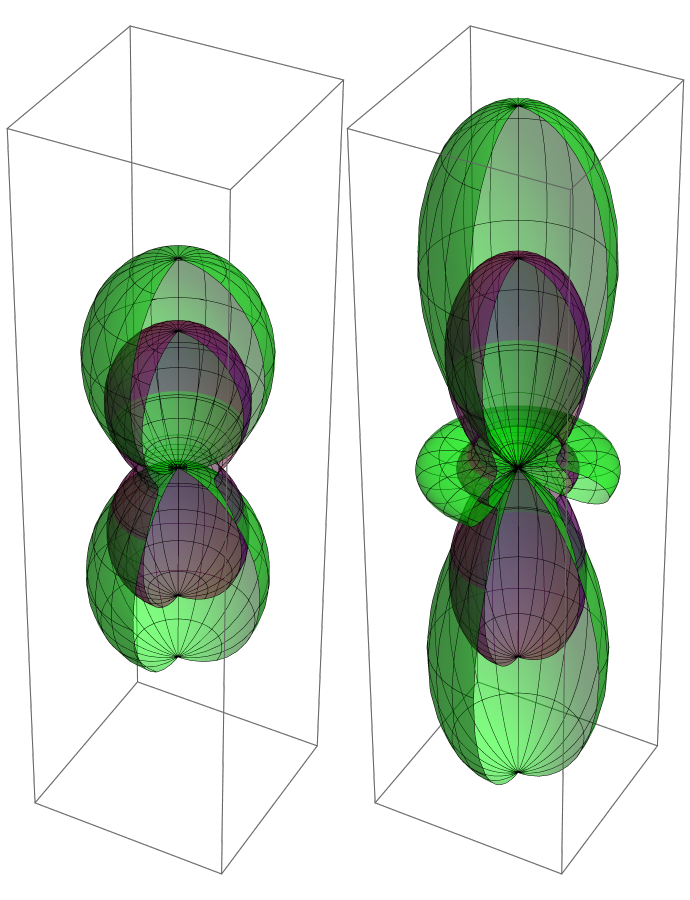}
\caption{\label{fig:orbitalcomp} The real spherical harmonics (green) $|Y_{10}|^2$ (left) and $|Y_{20}|^2$ (right) compared to the $|\mathcal{Y}_{3/2,high}^{\pm 1/2}|^2 / |\mathcal{Y}_{3/2,low}^{\pm 1/2}|^2$ (left) and $|\mathcal{Y}_{5/2,high}^{\pm 1/2}|^2 / |\mathcal{Y}_{5/2,low}^{\pm 1/2}|^2$ (right) Dirac spherical harmonics (purple). Note that while the Dirac spherical harmonics squared with the same $j$ and $j_3$ are the same, their radial functions are dependent on $high$ ($l=j-1/2=-\kappa-1$) and $low$ ($l=j+1/2=\kappa$), so their Dirac densitals will not be the same.}
\end{figure}

We will here show two more indications to support this fact. In Fig.~\ref{fig:orbitalcomp} we compare $p_z$ and $d_{z^2}$ squared orbitals for the real spherical harmonics with the corresponding $D^h p_{\pm 1/2}/D^l d_{\pm 1/2}$ and $D^h d_{\pm 1/2}/{D^l f_{\pm 1/2}}$ squared orbitals for the Dirac spherical harmonics. We note that the Dirac orbitals are contracted towards the center while still having \emph{more} probability density at the center. The Dirac orbitals are formed from superpositions of real orbitals so as to be better suited to accommodate a larger density around the nucleus. The orbitals for the $D^h s_{\pm 1/2}/D^l p_{\pm 1/2}$ are spheres and only the radial functions can contract these densitals.

\begin{figure*} [!ht]
\includegraphics[width=16cm]{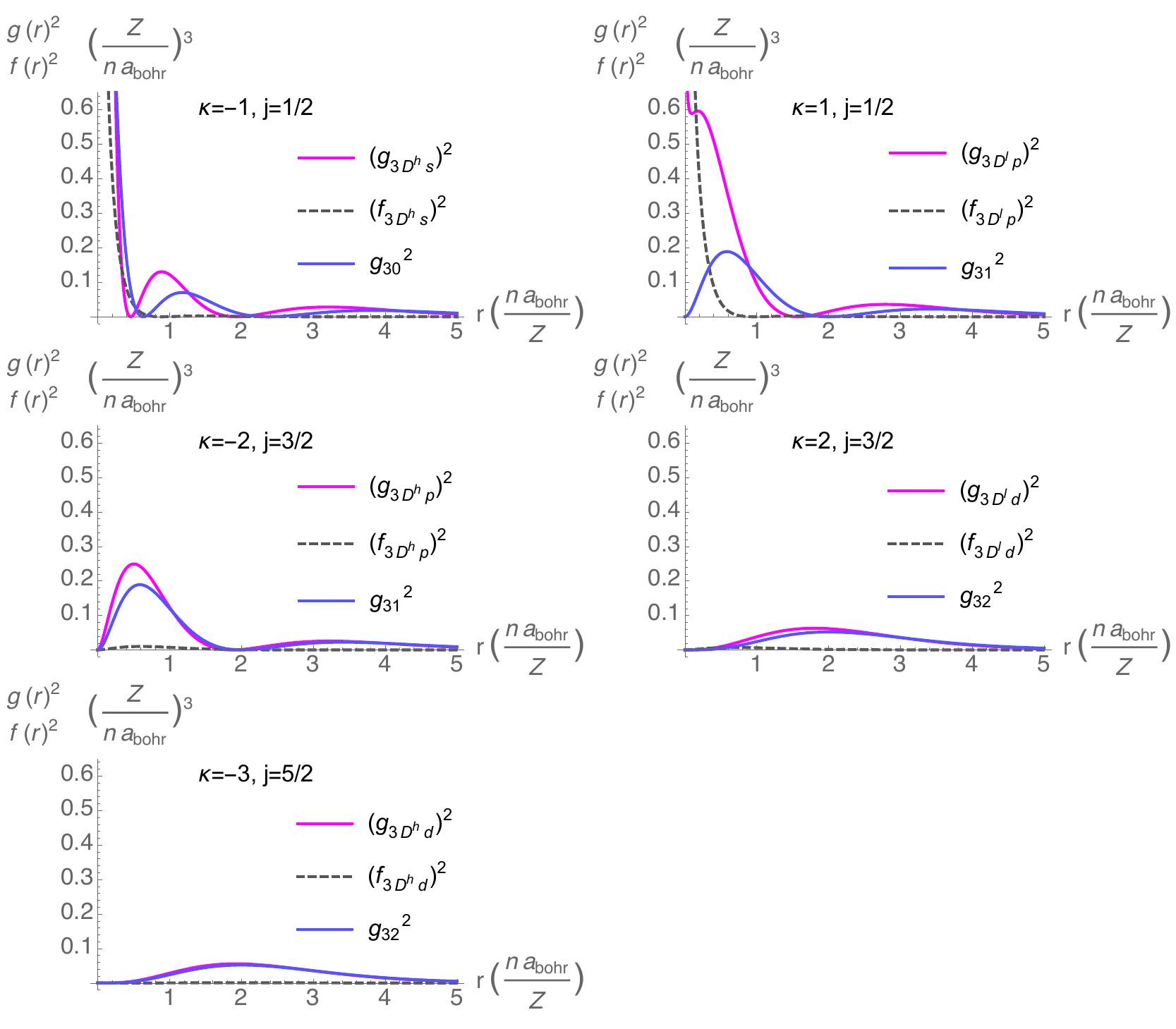}
\caption{\label{fig:n3radials} The five Dirac radial functions in shell $3$ compared to the three SE radial functions. Note that the angular part is determined by the value of $j$, so even if we use the SE radial functions instead of the Dirac radial functions, we will have 5 different groups of densitals in the third shell. Note also that the energy degeneracy for the Dirac solution only depends on the absolute value of $\kappa$, $|\kappa|$. While all third shell states are degenerate for the SE solution, each $l$ state will be split into a $high$ and $low$ energy state in the Dirac solution. This is the spin-orbit splitting.}
\end{figure*}
In Fig.~\ref{fig:n3radials} we compare the Dirac radial functions (for $Z=94$) with the non-relativistic ones. As we have noted before, while the squared orbitals for $D^h (l-1)_{j_3}$ and $D^l (l)_{j_3}$ are the same, their radial functions are not. We clearly see that the Dirac radial functions are shifted towards the origin compared to the non-relativistic ones. In addition we note that the lower-components Dirac radial function, $f_{N\kappa}$, have larger weight when the upper component Dirac radial function differ the most from its non-relativistic counterpart. The large difference between the $3p$ radial function and the $3D^l p$ radial function is particularly revealing. Finally we should note that the relativistic effects on the radial function is smaller for larger $l$. For larger $l$ the relativistic effects are almost exclusively in the orbital part of the solution. In summary, the contraction of the lower $l$ densitals are due to the radial function while the contraction of higher $l$ densitals are due to the orbitals. This means that if we use Dirac spherical harmonics also when evaluating the SE solutions, any relativistic effect should be included in the larger $l$ states.

The observations made in this section could explain the fact that the PBE~\cite{PBE} functional in DFT often gives relatively accurate structural predictions, i.e., atom positions and lattice constants, for actinide materials when studied using SE-based DFT despite these studies' non-relativistic nature.~\cite{soderlind2010assessing,kocevski2022uranium} If the true electron density in a relativistic material is more concentrated around the nuclei compared to the density in a non-relativistic calculation for that material, it means that the true interstitial density in a relativistic material is lower compared to the interstitial density in the non-relativistic calculation for that material since the total number of electrons must remain the same. The density overlap between atoms is thus smaller in the relativistic material compared to the non-relativistic calculation of that material. This means that bonds are weaker in a relativistic material than what a non-relativistic calculation (with a good functional) would give. Note that in the limit of no interstitial density we would have no bonds or only very weak van der Waals' bonds. It follows that the lattice constant can be expected to be larger in the relativistic material than what a non-relativistic calculation (with a good functional) would give. For non-relativistic materials, well treated by the SE, PBE has a tendency to underbind and give too large unit cell volumes. As described above, this is exactly the same tendency as the lower density in the interstitial region of a relativistic material would have. However, this means that, even though PBE in the SE treatment predicts accurate atomic positions and lattice constants, the density in the interstitial region is too high. This is commonly described as a \emph{functional} deficiency and not an effect of neglected relativistic effects. The common view is that the density becomes too delocalized because we have based the functionals on the fully delocalized  uniform electron gas. The remedy is often to introduce some localizing effect, such as a $+U$ treatment on the $f$- or $d$-orbitals. In view of the insights here, it might be that we should instead contract the $s$ and $p$ orbitals, which are the most contracted ones in the Dirac equation compared to the SE, and use a functional with correct binding properties on this contracted density. The key insight from this scenario would be that the regions in space that are accounting for binding even in relativistic materials have non-relativistic character, and all effects are secondary and due to the contraction of the density towards the nuclei. As is usually the case, the truth is probably that we both have to create new functionals able to describe both localized and delocalized electrons on the same footing, and use the full Dirac equation in order to obtain accurate and indisputable results for relativistic materials.

\subsection{Scalar relativistic method}
\begin{figure} 
\includegraphics[width=8.5cm]{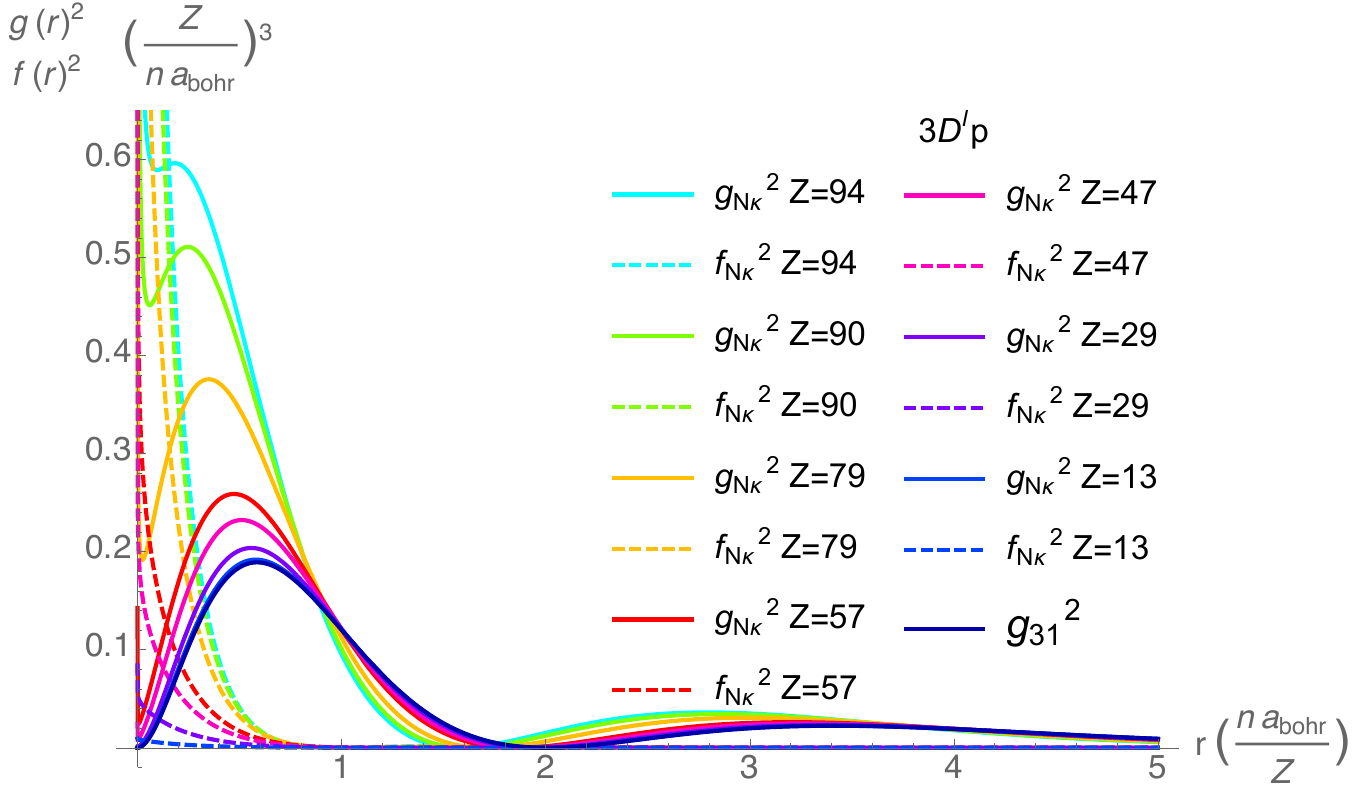}
\caption{\label{fig:DlpZtrend} Illustration of how the Dirac radial function approaches the non-relativistic radial function as the value of $Z$ is lowered. $Z=94$ corresponds to plutonium, $Z=90$ to thorium, $Z=79$ to gold, $Z=57$ to lanthanum, $Z=47$ to silver, $Z=29$ to copper, and $Z=13$ to aluminum.}
\end{figure}

The scalar relativistic method by Koelling and Harmon~\cite{koelling} is widely used and often is successfully extended to include spin-orbit coupling by a variational procedure on top of a scalar relativistic basis. The method is based on neglecting part of the angular momentum eigenvalue in the radial equations in order to obtain a radial equation only dependent on the non-relativistic quantum numbers $n$ and $l$. This radial equation, however, cannot be put in the same form as we have used for the Dirac and the SE above. In fact, even though the SR solutions might form a basis, this basis has the wrong boundary conditions compared to the Dirac equation and thus even a variational approach for including the omitted SO term will not give the full Dirac solution. This is a well known fact that effects the $D^l_p$ state the most. In Fig.~\ref{fig:DlpZtrend} we show this radial function for a number of different $Z$ values and how it successively deviates from the non-relativistic $3p$ radial function with growing $Z$. The behavior of the radial function near the nucleus is determined by the lowest power, $p-1$, of $\rho$ in the radial solutions above. For the Dirac solution $p-1=\sqrt{\kappa^2-\gamma^2} - 1$ and we see that both $\kappa=\pm 1$ gives an almost zero value. However, for the non-relativistic solution we have $p_{NR}-1 = l$ which is zero only for $s$ states. For the scalar relativistic solution the lowest power is $\sqrt{l(l+1)+1-\gamma^2}-1$ which, again, is zero only for $s$ states. 

We should note that the Dirac spherical harmonics are eigenfunctions of the spin-orbit coupling operator. This means that the spin-orbit coupling itself is included in any solution written in terms of Dirac spherical harmonics. In the non-relativistic limit, the coupling vanishes because the lower-component radial function becomes very small and we can rotate to a frame where spin and angular momentum are decoupled and each good quantum numbers. In the scalar relativistic method, however, part of the spin-orbit coupling effect on the radial function is explicitly omitted with catastrophic consequences for the $D^l p$ state.

\section{Conclusions and future work\label{sec:future}}
Just as the SE solution to the hydrogen-like atom is the basis for the $s$-, $p$-, $d$-, and $f$-electron language of non-relativistic electronic structure, we expect the Dirac solution of the same system presented here to form the basis for the language of relativistic electronic structure. We have provided a simple system for translating from the $l$, $m$, and $s$ non-relativistic quantum numbers to the $j$, $j_3$, and $\kappa$ relativistic quantum numbers via the $n \, D(irac)^{h(igh),l(ow)}l_{j_3}$ notation, with $n$ the non-relativistic principal quantum number and with $j=l+1/2$ for $D^h$ ($\kappa<0$) and $j=l-1/2$ for $D^l$ ($\kappa>0$). We have introduced Dirac spherical harmonics and densitals, which can be used for visualizing SE and Dirac solutions on the same footing. We have noted that the main effect of relativity in the case of a hydrogen-like atom is a contraction of the electron density towards the core and that this radial contraction is larger for lower angular momentum $l$, such as $s$ and $p$ states, while the larger angular momenta $d$ and $f$ orbitals is closer to the non-relativistic limit.

We intend to investigate the power of the here-presented alternative viewpoint in several lines of work. For one, it is often elucidating to examine the density of states (DOS) of a system. The integrated DOS is closely related to the partial charges in Eqn.~\ref{eqn:dens} that we based our definition of densitals on. To gain even more detailed information, an $lm$-decomposed DOS is often studied, in particular for actinide materials. Commonly the $lm$-decomposition is made by projecting on real spherical harmonics, but we want to explore what more information we can gain by instead projecting on Dirac spherical harmonics. Another line of work is to reexamine the scalar relativistic method from this alternative view point and see if we can modify it to resolve some of the most severe deficiencies. Basing all-electron methods and pseudo-potentials on the Dirac spherical harmonics is another obvious task. We expect this alternative viewpoint to help in developing existing and new computational methods to give better results for actinide materials.

\begin{acknowledgments}
This work was supported in part by Advanced Simulation and Computing, Physics and Engineering Models, at Los Alamos National Laboratory. Research presented in this article was also supported by the Laboratory Directed Research and Development program of Los Alamos National Laboratory under project numbers 20230518ECR and 20210001DR. This work was also performed, in part, at the Center for Integrated Nanotechnologies, an Office of Science User Facility operated for the U.S. Department of Energy (DOE) Office of Science. Los Alamos National Laboratory, an affirmative action equal opportunity employer, is managed by Triad National Security, LLC for the U.S. Department of Energy’s NNSA, under contract 89233218CNA000001.
\end{acknowledgments}

\bibliography{references}

\begin{thebibliography}{34}
\bibitem [{Dir()}]{Diracsphericalharmonics}%
  \BibitemOpen
  \href@noop {} {}\bibinfo {howpublished} {Other names we have seen for Dirac
  spherical harmonics are "Spinor spherical harmonics" and "Pauli central field
  spinors".}\BibitemShut {Stop}%
\end{thebibliography}

\end{document}